# Exploring new Approaches for Information Retrieval through Natural Language Processing

Manak Raj[1], Dr. Nidhi Mishra[2]
ASET, Amity University
Noida, Uttar Pradesh, India
manak.raj@s.amity.edu, nmishra1@amity.edu

**Abstract**

This review paper explores recent advancements and emerging approaches in Information Retrieval (IR) as applied to Natural Language Processing (NLP). IR is one of the domains under NLP, which is itself a vast field of study. The dynamic nature of IR has been propelled by significant technological innovations and evolving research paradigms. In the study of different research papers, information such as the evolution of IR models, advanced retrieval techniques, and their applications in various domains has been considered. Emphasis is placed on the methodologies used by different research scholars to enhance retrieval performance, particularly through the integration of techniques like reinforcement learning and deep learning. Different aspects of IR have been taken into consideration while designing this review paper. Some of these include private information retrieval, advancements in IR with models like BERT (Bidirectional Encoder Representations from Transformers) for improved retrieval accuracy, the functioning of search engines like Google, and the use of Lucene, an open-source library that facilitates efficient text indexing and searching. Additionally, the use of a Python toolkit called 'Pyserini' which supports reproducible IR research with both sparse and dense representations, is discussed. This toolkit exemplifies ongoing efforts to develop robust and flexible IR systems that cater to the complex requirements of modern information retrieval tasks. Other aspects of the topic are also included in this paper. It provides a comprehensive overview of the methodologies, applications, and performance of these approaches, highlighting the trends and future directions in the field. By integrating innovative techniques and addressing emerging challenges, the reviewed studies explain how the advancement of IR is ultimately enhancing the efficiency and effectiveness of information retrieval in various practical applications.

**Keywords:** NLP, Information Retrieval, Machine Learning, Deep Learning, Search Engines

# 1. Introduction

Information Retrieval (IR) and Natural Language Processing (NLP) are important areas in computer science, enabling the efficient extraction and utilization of information from vast datasets that will be worked upon. IR focuses on the retrieval of relevant information in response to user queries, while NLP involves the processing and understanding of human language by computers. This review examines new approaches in IR through NLP advancements, providing a comprehensive overview of contemporary models, techniques, and applications. Information Retrieval is one of the domains of NLP and these days has become a very important factor in user's experience. User will not want to have a poor response from the search engine like Google that is why it is very important that there is a continuous advancement in this field. NLP is the main center of knowledge of all the different domains like IR, Machine Translation, Semantic Analysis etc. Information Retrieval (IR) is a critical component of Natural Language Processing (NLP), dealing with the representation, storage, and access of information. With the exponential growth of digital data, efficient and accurate IR systems are essential. So, it is very important for us to understand what Natural Language Processing actually is. Natural Language Processing (NLP) is a field of study and practical application which is focused on enabling computers to understand and manipulate natural language (which is our Human Language text or speech) effectively. Researchers in NLP strive to understand how humans use language, and making it easier for many other fields of study to apply the principles in disciplines like computer science, linguistics, mathematics, electrical engineering, artificial intelligence, robotics, and psychology. By harnessing insights from these diverse fields, NLP aims to develop tools and methodologies that empower computer systems to interpret and manipulate natural languages for various tasks. Applications of NLP span a wide range of domains including MT, NLP processing and summarization, user interfaces, multilingual and cross-language information retrieval (CLIR), speech recognition (like Alexa, Google Assistant) and an artificial intelligence. As we now know what NLP is we can also strive to know what IR is. Information Retrieval. It is a field within computer science that deals with the representation, storage, organization, and retrieval of information items such as documents, web pages, images, audio, video and other types of data from the huge dataset. The goal of IR is to provide users with relevant information in response to their queries or information needs. Search engines like Google, Bing and more utilizes various IR techniques to index and retrieve information from vast amounts of data available on the web.

## 2. Literature Review

### 2.1 Information Retrieval Models

There are various IR models like Boolean Model, Vector Space model, Probabilistic model, Inference network model.**[3]**

a) Boolean Model

   The Boolean model, one of the earliest and oldest IR models, which represents documents as well as queries as clusters of expressions, employing Boolean algebra (AND, OR, NOT) for query formulation. Its incapacity to rank the documents affects its effectiveness for more complex queries. It has a retrieval function that sees a document as either relevant or irrelevant.

b) Vector Space Model

   In the Vector Space Model (VSM), all the documents and queries are portrayed as vectors in a multi-dimensional space.

   $$d_j = (w_{1,j}, w_{2,j}, \dots, w_{t,j})$$
   $$q = (w_{1,q}, w_{2,q}, \dots, w_{t,q})$$

   Figure 1: Document and query as vectors **[3]**

   The resemblance between a document and a query is found out using the concept of cosine similarity. VSM supports the ranking of documents based on coherence and relevance, which makes it quite popular in IR systems.

   $$sim(d_j, q) = \frac{d_j \cdot q}{\|d_j\| \cdot \|q\|} = \frac{\sum_{i=1}^{N} w_{i,j} w_{i,q}}{\sqrt{\sum_{i=1}^{N} w^2_{i,j}} \sqrt{\sum_{i=1}^{N} w^2_{i,q}}}$$

   Figure 2: Similarity Cosine function **[3]**

   VSM has one of its evaluation metrics which is a term which introduced term-weight scheme and known as TF-IDF weighting. All of these weights have a term frequency (TF) factor which helps in measuring the frequency of terms in the document or query. Whereas, inverse document frequency (IDF) factor is used to measure the inverse of the number of documents that contain a term either a document or a query one.

c) Probabilistic Model

The most crucial function of the probabilistic model is the concept of it to initiate ranking of documents on probability obtained from ability of relevance in any user's query. All the two, documents and user queries are represented by vectors ~d and ~q and these are known as binary vectors, each binary vector component shows whether a given document or term occurs in some document or query, or it does not.

d) Inference Network Model

In IRS(s), each and every document along with terms that has different degrees of impact. When a user submits a query, all these things from many terms are collected to calculate some numerical score for the documents. This can be viewed strength, as the weight of each term within the document, influencing its significance to the given query. Document ranking in these systems follows methods and techniques in models such as the vector space model, probabilistic models. Anyhow, the correct method for quantifying the power of a term within any document is not very rigid and can be accustomed according to different choices of design and aims.

## 2.2 Web Search Engines

Information Retrieval (IR) plays a critical role in the functioning of nearly all of the web search engines which we use a lot in our daily lives, they are very essential tools for finding the widened corpora of information which is available all over the internet. Web search engines use IR techniques so that it can index, search, and rank all the web pages in order to provide the users with relevant results which they were seeking and that is truly based on their queries given by the user. When any user gives a query, the search engine starts understanding it in order to identify the key terms present in there. The search engine then extracts a list of all the web pages from its database that has been indexed and that contain these long terms which are related to the user's intent. This process involves several steps, including:

a) Crawling - The crawler has a function of seeking and extracting the relevant documents for the given search engine like Google, Bing etc. Crawlers size and shape differs a lot, but one thing which is common to all the web crawlers is that it follows links on online pages to find the web page and then download all new pages. A web crawler has its own reach it can be limited to a place like an University.
b) Indexing – It is the second process after once crawling is done. It means that the system will now create a document index. In this phase, query is written by the user so that some relevant information can be retrieved.
c) Ranking – Once these two processes are done, the IR system searches the index for the relevant documents that are in accordance with the query and presents them to the user.

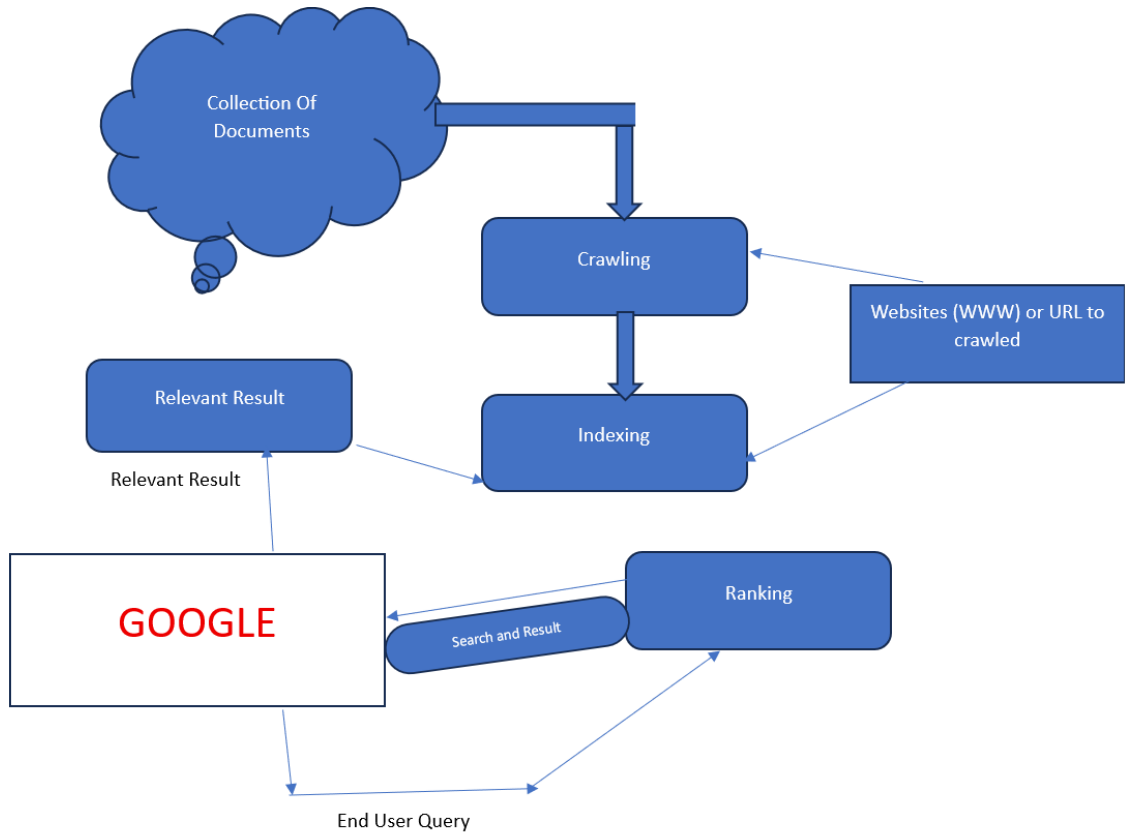

Figure 3: Web IR process [1]

### 2.3 CLIR Using BERT (Bidirectional Encoder Representations from Transformers)

Cross-Language Information Retrieval (CLIR) is a filed from information retrieval that primarily focuses on fetching all the documents that are written in a language that the user who is proving the query does not know. This is kind of tricky but it simply starts with the translation of queries into the document's language or the other way around. Use of MT, some multilingual sources, datasets cross this language barrier. The main purpose of all of this to allow the people access to information all across the globe without any barriers. Despite of all these innumerous problems, CLIR still plays an indispensable role in helping the world in global information access and enhancing the research of multilingual nuisances. BERT, XLNet and many more are the examples of the neural language models that have been developed recently and have achieved quite impressive results in nearly all of the NLP tasks including sentence classification, question answering and document ranking, information retrieval and many more tasks. BERT has been used in **[9]** for CLIR (Cross Lingual Information Retrieval).

According to the traditional norms, CLIR consists of two factors that play the roles here, the first one is machine translation and the next id monolingual IR. Cross-lingual IR has gained a lot importance with the advent of increase in the information available all over the internet. It is the first bidirectional language model and it uses right and left texts to predict the next texts. It is heavily trained by optimizing two aims one is masked prediction and another one is next sentence prediction. The self-attention mechanism which is present in all the BERT models, so it can learn pairwise sentence. In this paper, they have extended and applied BERT as a ranker for CLIR. They introduced a cross-lingual deep relevance matching model for CLIR based on BERT. They finetuned a pretrained multilingual model with their own CLIR data and obtained very promising results. Just to finetune the model, they made a large amount of data being trained from parallel data they had, which is mainly used for MT and is much easier to access as in comparison to the relevance of contexts.

## 2.4 Detailed Review of Selected Research Papers

| S No. | Name | Methodology | Result | Advantage | Disadvantage |
|---|---|---|---|---|---|
| 1. | Pyserini: A Python toolkit for reproducible information retrieval research with sparse and dense representations | The various methods that have been used in here are sparse, dense and hybrid retrieval. | Showed that hybrid retrieval overshadows the individual sparse or dense retrieval methods in terms of their retrieval effectiveness. | Easy to use, reproducibility factor, pre-built-indexes, high performance and support for all the three retrievals. | Pyserini relies mainly on several dependencies, like Java (in Lucene) and other Python libraries. Managing all these dependencies and ensuring compatibility can sometimes be challenging and not noteworthy. |
| 2. | Development of an IR System for Argument Search | It includes pre-processing, stoplist, stemmers, field weights and reranking have been used here. | BM25 initial nDCG@5 score was 0.3938. Where, LMDirichlet initial nDCG@5 score was 0.7345. This shows us the superiority of LMDirichet over BM25 | Improved the decision making of system, organization of content and improved research capabilities. | Intricacy of argument mining and scalability issues. |
| 3. | A Linguistic Study on Relevance Modeling in Information Retrieval | Model selection and finetuning, probing tasks (semantic, lexical, syntactic) and their performances have been used here. | The relative improvements of each and every retrieval models are 0.83%, 0.5% and 1.1% on document retrieval, answer retrieval and response retrieval. | In depth analysis, guidance for model selection and task specific insights for the system. | It has a very complex nature. |
| 4. | A simplified framework for zero-shot cross-modal sketch data retrieval | Encoder decoder architecture, cross triplet loss and zero shot learning algorithm have been used here. | It showed the effectiveness of the proposed approach. Important performance metrics, includes precision, recall, and F1-score, are significantly higher for the proposed method as compared to all the previous approaches. | Allows the retrieval of data without the need for every category, making it firm to various applications and new domains. | To develop a zero-shot model is quite complex and require some advanced understanding of both sketch data and cross-modal retrieval techniques. |
| 5. | An Information Retrieval Approach to Building Datasets for Hate Speech Detection | Selecting tweets via pooling, selecting tweets via AL(Active Learning) have been used here. | Detection on Twitter was created, yielding 14.6% relative coverage and 14.1% hateful content presence. | Reduces annotation costs through selection techniques like pooling and active learning. | Existing detection models struggle with the larger subtleties of hate speech covered by the new dataset. |
| 6. | Private Information Retrieval in Distributed Storage Systems Using an Arbitrary Linear code | System model, privacy model, and Communication Price Of Privacy (cPoP) have been used here. | Took the PIR protocol and where data is stored using an MDS code to the case where arbitrary systematic linear code is of rate R >1/2 and that is used to store data. | Linear codes in PIR protocols helped in achieving privacy. This is important for protecting user privacy. | The effectiveness of all the PIR protocols using linear codes often relies on certain security assumptions. If any of these assumptions are violated, the privacy gets hindered. |

## 3. Methodology

This methodology section's objective is to provide an overview of the main approaches used in some of the selected research papers. It contains the methodologies of the papers that have been chosen on the basis of their relevance to the field of IR and NLP. The traditional IR models such as BM25 and TF-IDF involve methodologies like ranking documents which are based on term frequency factor and inverse document frequency factor. For example, in the paper **[2]**, TF-IDF has been utilized to index and alongside rank the documents, getting some performance improvements. TF-IDF is a weighting method which is used to measure the importance of a word in any document among a collection of documents. TF is the term frequency which is the number of times a term q in document t. The IDF has a quite discriminating power. It is inverse document frequency which is used to measure how rare or frequent a term is in a collection of documents T. It gives high weight to lesser but frequent terms and a lower weight to more but frequent terms.

In the paper **[11],** SPLADE v2 has been discussed and SPLADE v2 incorporates three major methodologies:

1. Pooling Strategy: There is a change from sum to max pooling in the mechanism of term weighting that improves the performance of the model.
2. Document Encoder: It is a version of SPLADE that is totally replied upon expansion of document, enabling the pre-computation of term weights which eventually reduces inference costs.
3. Distillation Training: It is a two-step process where firstly the model is trained using triplets which have been generated by the previous models and after that it is again trained with harder negatives which are generated by the SPLADE before the process of distillation.

In the paper **[20]**, authors have used two methodologies:

**Reinforcement Learning Algorithm**: The methodology in this study involves adding up a reinforcement learning algorithm into the LLaMA model. The RL algorithm is designed to adapt the model's response in accordance with user queries, giving a combination of dynamical systems theory and relativistic physics to model the temporal arise of IR strategies. This system uses reward-based learning to refine retrieval parameters like it on loop, increasing the model's accuracy, relevance in IR tasks. The algorithm's decision-making process is due to this complex transformation:

$$c_i = \begin{cases} \sum_{j=1}^{N} \left[ \Theta\left(\gamma - \frac{\partial}{\partial t} \int_{\Omega} p_j(t,x)\, dx\right) \right], & \text{if } e_i = 0, \\ \sum_{k=1}^{M} \left[ \Lambda_k \left(V_i, R_i, \vec{\nabla} \cdot \left(\mathbf{D}\vec{\nabla} p - \mathbf{u} p\right) + \kappa p\right) \right], & \text{if } e_i = 1, \end{cases}$$

Figure 4 : RL Algorithm **[20]**

Where, $c_i$: Control action for retrieval task, $e_i$: Error in retrieval, $\Theta$: Heaviside step function, $\gamma$: Threshold parameter, $\Lambda_k$: Lorentz-like transformation function, $V_i$: Value function, $R_i$: Reward threshold, $\kappa$: Scaling factor

**LLaMA Model Adaptation**: Adapting the LLaMA model involves some implementation of the RL algorithm within its pre-existing architecture. The process is structured to adjust the retrieval parameters based on user's feedback, thereby optimizing the model's responses to align more closely with user queries. It aims to reduce the common issues such as irrelevant IR and hallucinations, which are typical in baseline LLMs.

The proposed methodology in the paper **[10]** integrates several steps to enhance IR performance using social data. Key components include:

1. Query Expansion: One more time, processing the user queries by integrating more relevant terms derived from the content and social tokens of the documents.

2. User Scoring: Finding the user's intent based on their interactions and contributions. The scoring system under this helps to combine all the measures of user.

3. Document Weighting: Assigning weights to documents based on user scores and interactions, enhancing the relevance calculation of documents to queries.

4. Performance Metrics: The use of precision and recall that helps to find the effectiveness of the proposed approach as in comparison to standard IR methods.

## 4. Conclusion And Discussion

There has been a number of significant advancements in this dynamic field, which is information retrieval through natural language processing. If we see all the recent developments that have taken place in the area of machine learning, particularly deep learning and reinforcement learning, they have transformed all the traditional IR models into more refined systems that are capable of understanding and processing human language to an impeccable level. The summation of all the models like BERT, COLBERT and so many other models, it has helped in increasing the retrieval accuracy and it makes the information retrieval systems work way more efficiently. We also have tools like Pyserini and Anserini that are used for reproducible information retrieval research and they are both used as they support sparse and dense representations that also enhances the capabilities of IR systems. All these advancements have come through the continuous efforts by the research scholars to meet all the difficult requirements of the modern era.

The improvements in IR through NLP have not only improved the efficiency of this field but has also given a hope for the expansion of this field to further domains and different applications. The adaptation of 'Entity Recognition', 'Semantic Analysis', although they are not the primary focus of this review paper, but still all these enhances the ability of IR systems and make them able to deal with the complex queries and the user interactions. However, these advancements come with many challenges and limitations and research scholars are still working for making this field more refined.